# Measuring Research Interest Similarity with Transition Probabilities

[PREPRINT]

Attila Varga, *Corvinus University of Budapest*

Sadamori Kojaku, *Binghamton University*

Filipi Nascimento Silva, *Indiana University Bloomington*

**Abstract**

We introduce a family of paper and author similarity measures based on the concept that papers are more similar if they are more likely to be retrieved during a literature search following backward and forward citations. Since this browsing process resembles a walk in a citation network, we operationalize the concept using the transition probability (TP) of random walkers. The proposed measures are continuous, symmetric, and can be implemented on any citation network. We conduct validation tests of the TP concept and other extant alternatives to gauge which metric can classify papers and predict future co-authors most consistently across different scales of analysis (co-authorships, journals, and disciplines). Our results show that the proposed basic TP measure outperforms alternative metrics such as personalized PageRank and the Node2vec machine-learning technique in classification tasks at various scales. Additionally, we discuss how publication-level data can be leveraged to approximate the research interest similarity of individual scientists. This paper is accompanied by a Python package that implements all the tested metrics.

## 1. Introduction

Understanding how specialized knowledge anchors researchers' attention, shapes boundaries, and enables interactions is a fundamental issue across various fields of science and technology studies (Gieryn, 1999). This paper focuses on a fundamental dimension of scholarly communication: research interest similarity. This concept is relevant for a wide range of issues, including technical challenges such as field normalization in scientometric indicators (Ioannidis, 2016; Hicks et al., 2015) and the measurement of interdisciplinarity (e.g., Stirling, 2007; Wang & Schneider, 2020; Cantone, 2024). It is also central to substantive research, such as global inequalities in attention allocation among researchers (Gomez et al., 2022) and the evolution of problem choice in researcher careers (Gieryn, 1978; Zeng et al., 2019; Yu et al., 2021).

The proposed measure of similarity is based on a model of information seeking behavior of scientists interacting with the literature. Two papers are considered more similar if they are more likely to be retrieved during a literature search following backward and forward citations. This paper-level measure can be aggregated to higher levels, such as authors, fields, or institutions. It can also be utilized to examine the similarities of paper's references and to calculate paper-level indices. Among these possible aggregates, this paper focuses on scientist-level similarity, as the definition is inherently tied to authors and their efforts to update their knowledge. Additionally, we consider researchers' problem selection a pressing issue and aim to contribute conceptually and methodologically by sharing our experience in aggregating paper-level data to represent authors' research interests.

We introduce a family of indices based on random walk transition probabilities (TP). These measures have several key properties. First, we build the concept on an explicit definition of similarity by following a simple information behavior model, providing an interpretable metric. Our approach is based on an explicit definition of similarity derived from a simple model of information behavior, making it an intuitive and interpretable metric. Since the construct is rooted in observable behavior, it can be empirically studied (Wilson, 1999). This also aligns with theoretical arguments and agent-based models referencing researchers' attributes and motivations (Gieryn, 1978; Rzhetsky et al., 2015). These theoretical and empirical insights can, in turn, refine the measure to incorporate factors such as agent biases or new information-update rules. Second, our measure is fundamentally symmetric, which is a desirable quality for similarity measurement. Third, we introduce continuous similarity measures, which we argue are, in certain respects, more suitable than discrete representations for tracing the cognitive organization of scientific progress. Fourth, our measure maintains more consistent precision across different analytical scales compared to related methods. To test the scale consistency and the overall validity we conduct three experiments at different levels of aggregations: collaborations, journals, and disciplines.

In section 2 we review past approaches and relevant computational methods to research interest similarity. Section 3 defines the concept of browsing distance and its calculation based on TPs of random walks. Section 4 introduces alternatives measures. These are either computationally efficient substitutes of the original basic TP metric, or similar well-known or trivial alternatives. After describing the data in section 5, section 6 outlines the three validation tests for the metrics. The results of these experiments and the comparison of calculation runtimes for each method are presented in section 7. This section also evaluates the correlation of the metrics with the nodal degree. Section 8 summarizes our key findings and insights about measuring research similarity at the author level with publication data. The paper also briefly presents (in Appendix C) a Python package to calculate all the indices discussed in this paper.

## 2. Related Research

Framing the stages and interactions of evolving research agendas is often done in terms of discrete topics or problems (e.g. Gieryn, 1978; Rzhetsky et al., 2015; Jia et al., 2017). However, we argue that when the similarity of research programs is the central concern, continuous representation is more justifiable than clustering evolving research into discrete problems. While well-defined research problems can be constructed for research administration purposes or by hindsight after a research program is ended, it is more realistic to compare two scientists' research on a gradual scale, rather than based on a rigid discrete categorization.

Previous theoretical work on research interests has used the concept of research problems to explore why and under what circumstances scientists change their research focus. (Gieryn, 1978). Although the term "research problem" suggests clear-cut objectives, even this line of inquiry acknowledges the difficulty of breaking research activities into well-circumscribed problems. Gieryn (1978) in his work on problem retention and change in science discussed the span of active research problems of a scientist at a given moment, and how these interests evolve over time. According to Gieryn, scientists usually work on a unique selection of interrelated problems. When they transition to new areas their new problem set remains influenced by and mixed with their previous work. Given that a research agenda is defined as a structure of multiple interrelated problems, a scientist's shifts in attention tend to be gradual. Akin to this observation Ziman (1987), while criticizing the concept of problem choice, notes that research objectives constantly shift as work progresses. Scientific research is highly uncertain, and initial formulations of research problems can radically evolve over time. More recently, Cooper (2009) also emphasized that research problems are not clearly demarcated in ongoing research.

In recent years, research on authors has gained prominence as more name-disambiguated data has become available. Current approaches to modeling researcher attention often rely on discretized representations of topics. These topics may originate from a thesaurus or a collection of keywords (Rzhetsky et al., 2015; Leahey et al., 2017; Jia et al., 2017; Yu et al., 2021), clustering of bibliographic couplings (Zeng et al., 2019), or topic models utilizing textual information (Rosen-Zvi et al., 2012; Amjad et al., 2018). Aside from the theoretical issue of discretizing research agendas, these approaches have several practical drawbacks. Well-curated, up-to-date classifications of research problems are often unavailable, or when they do exist, they may not be at the appropriate level of resolution. While research specialties and problems can be mapped using clustering methods, clustering involves several subjective choices that affect the final output (e.g., determining the optimal clustering resolution, filtering out smaller clusters). Additionally, the distribution of clusters is often heavily skewed, requiring careful calibration to ensure that the chosen clustering method maintains a consistent level of resolution across all clusters.

Several techniques provide continuous measures of similarity, which can be applied to bibliometric data to analyze researchers and their evolving interests. Continuous representations encode the structure of topics in a spatial embedding, where papers are positioned close to other papers on similar subjects. These embeddings can be constructed from textual content (Cohan et al., 2020; Constantino et al., 2025) or citation networks (Tang et al., 2015; Kojaku et al., 2021; Peng et al., 2021; Choi and Yoon, 2022; Constantino et al., 2025). Traditionally, continuous representations have been generated using dimensionality reduction techniques applied to keyword co-occurrence matrices (Ahlgren and Colliander, 2009) or network adjacency matrices (Kunegis and Lommatzsch, 2009; Qui et al., 2018). More recently, a shift toward neural embeddings – generated by neural networks – has gained momentum (Perozzi et al., 2014; Grover and Leskovec, 2016; Hamilton et al., 2017; Kipf and Welling, 2017). These spatial representations overcome many of the limitations of discrete methods and offer numerous analytical advantages. However, neural embeddings present a critical challenge: due to the opaque nature of neural networks, they lack interpretability. It is often unclear what the similarity in the generated space represents or how meaningful insights can be derived from it. This lack of interpretability limits their usefulness in investigations into how knowledge structures anchor researchers' attention.

The proposed method is based on random walks on networks. As a fundamental network concept (Norris, 1997), random walks have been widely used in network analysis and have played a central role in various applications. These applications include for example measuring node importance (Page et al., 1999; Haveliwala, 2002; Jeh and Widom, 2003), link prediction (Liu and Lü, 2010), and detecting network communities (Pons and Latapy, 2005; Rosvall and Bergstrom, 2008; Delvenne et al., 2010). Among these applications the closest to our work is recommender systems (Baluja et al., 2008; Ren et al., 2014; Shi et al., 2014). In this context, random walks are typically one component within more complex algorithms. However, aside from the technical format, recommender systems differ from our approach in one key respects. Recommender systems are optimized to retrieve the most similar entities, whereas our approach is designed to measure the full spectrum of similarity – that is, to assess the similarity even between highly dissimilar items.

A highly relevant approach for measuring node similarities in networks, which is also based on the concept of random walk TP, is personalized PageRank (PPR; Haveliwala, 2002; Jeh and Widom, 2003). PPR extends the PageRank algorithm to capture node-specific importance scores by teleporting the walker back to its starting node. While PPR has been successfully applied in recommendation tasks (Bagci and Karagoz, 2016; Jiang et al. 2018), it is less suitable to measure research interest similarity for several reasons. First, this TP measure is suitable to retrieve the most similar nodes in comparison to the input node, making it less effective at assessing similarities between more distantly related nodes. In other words, it is designed to distinguish highly similar entities, but it is not sensitive to low-level similarities. Second, PPR produces asymmetric similarity scores (the similarity from node A to B differs from B to A), which contradicts the intuition that similarity should be symmetric. Third, exact computation of PPR is expensive for large networks. While approximate methods exist (e.g. Wu et al., 2021), they often trade-off accuracy for computational efficiency. This trade-off arises because PPR lacks a cutoff to limit random walks, meaning that, theoretically, it can explore the network indefinitely, resulting in a fully dense similarity matrix.

Recently, recognizing the limitations of bibliographic coupling (BC) and co-citation (CC) as paper similarity measures (due to the sparsity of such links in citation networks), Yun (2022) tested the efficiency of PPR and Node2vec embeddings (N2V; Perozzi et al., 2014; Grover and Leskovec, 2016) on node split networks (Yun et al., 2020) as an alternative approach. This work is highly relevant to our study, as we also investigate the same similarity measures within a bibliometric context. Yun (2022) found that many highly similar paper links could not be detected by bibliographic coupling and co-citation methods. Our study further highlights the inadequacy of BC and CC as similarity measures and shows that N2V and PPR have low correlation, agreeing mostly at high similarity ranges.

## 3. Concepts and Calculation

The concept of research interest similarity is based on how scientists learn about new literature related to their research agenda. A typical scientist would begin with a set of papers they are already familiar with. By following the forward or backward references of these papers, they would discover new papers. This process would then continue iteratively, as the scientist explores the references or citations of the newly found papers, until they acquire the desired new information about the current state of the art. In information retrieval, this search procedure is often referred to as pearl-growing search or snowballing search. It is an effective method for retrieving complex information (Greenhalgh & Peacock, 2005) and is likely familiar to most researchers. Based on this form of information retrieval, we define our similarity measure as follows: two papers are more similar if during a literature search starting from one of the papers it is more likely to retrieve the other paper.

We do not claim that this is the primary way researchers discover new and relevant papers. However, we assume that this process, along with other mechanisms serving the same function (such as attending conferences, discussions with colleagues, and social media interactions), would yield similar results. Therefore, we consider this definition to be sufficiently general. It is important to note that we are not proposing a new information retrieval model. Rather, we are drawing upon an existing concept in information retrieval and assuming that it is most likely a typical information seeking behavior in the sciences.

Our operational definition of similarity is based on the observation that this type of literature search resembles a walk in a citation network. Therefore, a scientist's search behavior can be modeled as a random walk. The operational definition of similarity is the following: the similarity of two papers $i$ and $j$ is proportional to the probability that a fixed-length random walk from $i$ passes through $j$. With this definition, we introduce a generalizable similarity measure for graphs, which, in this context, applies to academic papers. The research interest similarity of authors or the similarity of scientific fields can then be derived by aggregating the similarities across sets of papers, a topic we explore in more detail in the following sections. At this point, we emphasize that our approach provides a concise conceptual framework for understanding paper similarity. Unlike previous methods, our framework is based on an explicit mechanism of information update, which is rooted in purposeful action and shaped by the enabling/constraining information structure, encoded as a citation network.

The specific measure implemented here is the probability that a random walk starting from node $i$ transitions through a target node $j$ at any step during the walk in $t$ steps. The following definition assumes that the network is connected and undirected, and these assumptions are consistent with the conceptual definition above. We start the definition with $P_{ij}^{\tau}$, which is the probability that a random walk from node $i$ is at node $j$ after $\tau$ steps. This TP can be calculated by summing up the probabilities of every possible $\tau$-step long random walk, starting from $i$ and terminating at $j$. The TP of a single random walk is given by multiplying the reciprocals of each node's degree along the walk from $i$ to $j$, but NOT including the terminal node $j$. Finally, summing up all $P_{ij}^{\tau}$ probabilities from 1 to $t$ ($\{1, \ldots, \tau \ldots, t\}$) gives the probability that a random walk from $i$ crosses $j$.

This probability is not symmetric $\left(P_{ij}^{\tau} \neq P_{ji}^{\tau}\right)$, because if let's say, $j$ has a higher degree than $i$ ($k_i < k_j$), $j$ would be visited more often by random walks than $i$. We symmetrize the $P_{ij}^{\tau}$ probabilities by dividing them with $j$'s degree: $P_{ij}^{\tau}/k_j$. This essentially means that the computation of a single walk's TP changes to multiplying together ALL nodes' degree reciprocals, including $j$ as well. This operation is an addition to the conceptual definition. The rationale for this step, aside from simplifying the measure by symmetrizing it, is that the expected similarity of a paper should not be higher just because it has more references and/or citations. Although it is true that higher impact papers, or papers with more references are retrieved more often in a search, for our purposes it is unwanted to assign higher average similarities for highly cited papers.

The final TP, which is our basic measure $T_{ij}^{t}$ can be computed by using matrix operations in the following way. Denoted by $A$ the adjacency matrix of the citation network, where $A_{ij} = 1$ and $A_{ij} = 0$ represents that a paper $i$ is connected by an edge with paper $j$, or not, respectively. Recall that we ignore the direction of the citations and thus $A_{ij} = A_{ji}$. We use the transition matrix $P$ of random walks, where the TP of a walker from paper $i$ to paper $j$ is given by $P_{ij} = A_{ij}/k_j$, or equivalently, $P = D^{-1}A$ , were $D$ is a diagonal matrix, with diagonal entries being the degree of nodes, i.e., $D_{ii} = k_i$. The TP that a walker from node $i$ would make the $\tau + 1$-th step to node $j$ after $\tau$ steps can be found at the $(i, j)$-th entry of the $\tau + 1$-th power of the matrix $P$: $(P^{\tau+1})_{ij}$. Finally, $T_{ij}^t$ is given by adding up each $\tau$-th powers of the transition matrix from 1 to $t$, and applying the appropriate normalizations:

$$T_{ij}^t = \frac{1}{t}\left(\sum_{\tau=1}^{t} P^{\tau} D^{-1}\right)_{ij}$$

The division with $t$ assures that the final value correctly expresses the TP of a single step along an $t$ length walk, in other words it ensures that it $T_{ij}^t$ is bound by 1.

## 4. Alternative Formulations

Besides the above defined basic metric $T_{ij}^t$, we considered six alternative indices. The first group consists of measures derived from our original concept, designed to address specific shortcomings of $T_{ij}^t$. The second group includes well-known measures that may appear conceptually similar to our approach. By incorporating this latter group, we aim to highlight the specific advantages of our method.

Calculating the proposed TP is computationally demanding. The TP is based on the powers of $P$ that may hold $NN$ transition probabilities, where $N$ is the number of nodes in the network. The number of transition probabilities increases quadratically with respect to $N$, and thus, it is practically challenging to hold them in memory for large networks. For this reason, it is worth exploring alternative measures of $T_{ij}^t$. We considered two such alternatives, which are conceptually similar to the above defined TP, and would be reasonable substitutes of the original formulation.

As the first alternative, we propose a scalable method to estimate the $T_{ij}^t$ values by simulating random walkers in the networks. Namely, we launch $n$ number of random walkers from each node $i$, and all the walkers make $t$ steps. The crossing frequency of these walks on node $j$ is $x_{ij}^t$. The estimated transition probability is:

$$ET_{ij}^t = \frac{x_{ij}^t}{ntk_j}.$$

While the matrix computation necessarily gives the $T_{ij}^t$ of all the node pairs in the network, the advantage of $ET_{ij}^t$ is that it enables the sampling of node pairs. However, the estimation approach has some caveats. $ET_{ij}^t$ can be zero if there is no path with length $t$ between the nodes (in which case $T_{ij}^t$ is also zero), or if $T_{ij}^t$ is so small that none of the random walkers were able to span the network structure between node $i$ and $j$ during the estimation procedure. This essentially means that the estimation has failed. One can increase the number of walkers to obtain an estimate, but it's important to bear in mind that this approach exhibits diminishing returns. The chance that at least one walker will cross and give an estimated value rise at a decreasing rate with the number of walkers delegated to this task. Also note that $ET_{ij}^t$ is only approximately symmetric. Assuming $k_i > k_j$ simulated random walks from $i$ are more likely to give an estimate for $ET_{ij}^t$ instead of remaining zero. This is because higher degree starting nodes of random walks spread the walks more evenly in the network, and these walks are more likely to reach distant nodes. Therefore, when we calculate $ET_{ij}^t$ we simulate walks from both directions and take the average.

Another way to reduce the proportion of zeros estimates and gain better precision for low similarity ranges is to increase $t$. Increasing $t$ can aid the walker to span distant nodes with low $T_{ij}^t$. The choice for walk length $t$ can help to balance out the problem of reachability, and the fact that with the increase of $t$, $P^t$ is converging to a stationary distribution, which is proportional to the degree of the nodes, and would not provide meaningful information about similarity. We will return briefly to the issue of choosing the value of $t$ in the concluding section.

The second substitute introduced here, which still adheres to the behavioral model, is a crossover of TP and shortest paths (SP), and it is referred to as the average shortest path transition probability ($ST_{ij}^t$). For this measure one has to find all the shortest paths between two nodes $i$ and $j$, calculate the transition probability $p_{ij}$ of each shortest path $p_{ij} = \prod_{\tau=1}^{t} \frac{1}{k_\tau}$, where $k_\tau$ is the degree of the $\tau$-th node on the shortest path, and average out these $p_{ij}$ probabilities. This is a symmetric quantity, however conceptually it detours from the original definition. One can interpret this as a knowledgeable agent's search, whose browsing behavior is informed or biased to find the most direct connection between papers $i$ and $j$, and narrowing the search to corpuses of related literature such as by considering the paper's content.

The first and most basic alternative of our methodological approach is the shortest path length. It is the length of any shortest path between two nodes and can be computed by most network analysis programs. The second alternative measure is derived from the traditional science mapping tools BC and CC. It is the Jaccard coefficient of the two papers neighborhood in the citation network (denoted as $N(i)$), which includes all forward and backward references:

$$JAC = \frac{|N(i) \cap N(j)|}{|N(i) \cup N(j)|}$$

It expresses the level overlap between the two papers' forward and backward references.

We considered the N2V algorithm as an alternative of $T_{ij}^{t}$. N2V is a technique to embed nodes of a network into a vector space, generating a vector for each node. The embedding is based on a neural network trained on the sequence of nodes visited by random walks. It has been shown that the embedding generated by N2V is equivalent to the eigenvectors of the matrix $T^t = \frac{1}{t}\sum_{\tau=1}^{t} P^\tau D^{-1}$ (Qui et al., 2018), which relates to our measure.

Finally, we tested PPR. As was mentioned in section 2, PPR is a widely used similarity measure that is conceptually similar to our approach, as it is also based on random walk TPs. However, PPR is not well-suited for measuring low-range similarity, due to the restart mechanism in TP calculation. At each step of a random walk, the algorithm has a teleportation probability $p$ that allows it to return to the source node. Since the first step is always taken, the probability of a walk of length $l$ (where $l > 1$) occurring decreases exponentially with $(1-p)^l$. In simple terms, PPR prioritizes exploring the immediate vicinity of the focal node, whereas $T_{ij}^{t}$ does not exhibit this bias. To compute PPR efficiently, we implemented the Power Iteration with Forward Push method (Wu et al., 2021).

## 5. Data

The citation data is sourced from the Web of Science Science Citation Index (SCI), while disambiguated author information comes from the Microsoft Academic Graph (MAG), which has been matched to papers indexed in SCI. Using DOIs and bibliographic metadata, we successfully matched 90% of the papers to MAG. Research field information is derived from Web of Science subject categories (SC). We relied on MAG for author names, as Web of Science does not provide disambiguated author identities.

For our tests, we selected three research fields plus a multidisciplinary journal set. Each field is analyzed for a focal year between 1980 and 2019. The multidisciplinary set includes all articles from *Nature* and *Science*. Table 1 summarizes the selected fields and focal years. This selection ensures disciplinary diversity and spans different time periods, enhancing the validity of the results. The choice of fields and specific years was primarily constrained by computational feasibility, as calculating $T_{ij}^t$ is resource intensive.

Our goal was to create a diverse dataset with manageable citation networks, allowing for efficient $T_{ij}^t$ computation. The citation networks consist of 20-year slices of SCI-indexed papers, where the last year of each window is the focal year, and the first year is 20 years prior. For each field, the initial selection includes all articles from relevant journals within the given period. Those papers that don't have at least three connections in the network (considering all citations and references) are removed, and after that step the main component is selected for the analysis.

| Subject Category | Abbrev. | Year | Nodes | Edges |
|---|---|---|---|---|
| Astronomy & Astrophysics | Astro | 1980 | 65,535 | 983,206 |
| Clinical Neurology | Neurol | 1990 | 78,159 | 801,260 |
| Sociology | Soc | 2010 | 46,832 | 398,512 |
| Multidisciplinary | Multi | 2019 | 51,942 | 356,930 |

***Table 1***. *Summary table of datasets*

## 6. Validation Methods

The testing scenarios are designed to evaluate the performance of the similarity metrics across diverse use cases. Our primary concern is whether these metrics can consistently measure similarity at different scales and levels of aggregation. To devise these tests, we leverage the hierarchical nature of scientific research, where specific research goals are nested within specialties, which in turn belong to broader disciplines. These levels of scientific organization serve as ground truth for our validation tests.

Before detailing the validation tests, we briefly address why a consistent measurement scale is particularly important in science studies, whereas it is less critical in recommender systems and information retrieval. Estimating and distinguishing similarity at the mid- and macro-scale is crucial in several research contexts, particularly when studying information diffusion beyond local and direct interactions (Nakamura et al., 2011; Varga, 2019). It is also relevant in interdisciplinarity research, where long-distance interactions play a significant role (Larivière et al., 2015). Furthermore, it is essential for studies examining the relationship between cognitive structures and institutional processes within broader disciplinary organizations and policies (Gieryn, 1999; Becher & Towler, 2001; Mohr & White, 2008). Scientists frequently engage in collective decision-making where research interest similarities are low but still influential in forming a collective identity. Such scenarios commonly arise in faculty hiring, funding allocation, and large-scale data collection efforts (Brooks et al., 2016).

To evaluate the validity and precision of the similarity metrics, we designed three levels of validation tests. At the micro-level of co-authorship evolution, we tested how well a given method predicts future first-time collaborations based on past research similarity. At the mid-level of research specialties – approximated by journals – we measured how accurately the method distinguish between papers published in the same journal versus papers from different journals within the same discipline. Finally, at macro-level the test was to classify papers based on disciplinary background. Our expectation is that N2V will better capture local relationships but provide a coarser representation at the mid-to-macro range, as N2V is known to be less accurate for weaker similarity ranges (Grover & Leskovec, 2016). In all three cases, the test data consists of two variables. First, a binary variable indicating whether (a) authors have collaborated, (b) papers are in the same journal, or (c) papers belong to the same discipline. The second variable is a predictor/classifier variable, which is the tested similarity metric.

In the following we will describe the collaboration prediction task in detail. This scenario also introduces practical methods for measuring author-pair similarity. The collaboration prediction test consists of two sets of author pairs. Future collaborators, authors who have never collaborated before but will collaborate in the year following the focal year, and non-collaborators, randomly matched authors who did not collaborate before and will not collaborate in the following year. All selected authors are actively publishing and collaborating in the following three years after the focal year with someone. These selection criteria are necessary to keep the two sets comparable. Both the collaborator and non-collaborator author pair sets are based on a sample of 500 authors, except in sociology (232 authors) and multidisciplinary journals (431 authors) due to data constraints. Each author is paired with one of their future collaborators for the true collaboration set and with a random author for the non-collaborator set. We use a 5-year window to select papers written by the focal author, which serves as the basis for computing potential collaborator similarity. This same window is used to determine whether authors have previously collaborated. To measure research interest similarity, we consider only papers where the focal author is listed as the first or last author, assuming these roles better reflect their research interests. Our experience suggests that this selection improves prediction accuracy. The final author similarity score is the average pairwise similarity of their selected papers.

For the journal classification task, every journal within the given discipline was first selected for analysis. Then, 20 papers were randomly chosen from each journal. In the case of multidisciplinary data, where only two journals were present, 500 papers were selected from each. Each paper was then matched with one from the same journal and another from a randomly chosen journal. Journals that did not have a sufficient number of papers within the given period to meet these sampling criteria were excluded from the analysis.

For the macro-range test, we classified the subject categories (SCs) and their corresponding journals into six broadly defined disciplines: environmental sciences, agriculture, medicine, biology, chemistry, and physics. This test was conducted exclusively on multidisciplinary journal data (papers from *Nature* and *Science*). A paper's disciplinary background was determined by identifying the most frequently occurring discipline among its references. The sample consisted of 10,000 paper pairs, with half belonging to the same discipline and the other half to different disciplines. To ensure balanced representation, all disciplines were given equal weight in the dataset.

## 7. Results

We begin with descriptive statistics. These exploratory results are based on random samples of dyads with a size of 2000 paper-pairs in each field. Table 2 lists the tested measures and their abbreviations, while Appendix A presents the parameterization of the measures. The proportion of paper dyads with at least one overlapping citation or reference is less than one percent in all cases. Therefore, we do not present descriptive and bivariate statistics for JAC in the following. Figure 1 displays the distribution of the measures across the four dyad samples. The distribution of the TP measures roughly follows a lognormal distribution (Fig. 1/a and Fig. 1/d). The exception to this lognormal shape is Astronomy & Astrophysics, which exhibits a bimodal lognormal distribution. We attribute this uniqueness to the discipline's pronounced dual organization into two subfields: cosmology and observational astronomy (Varga, 2018). Within-subfield paper pairs are similar to each other, while between-subfield pairs are dissimilar, creating two distinct peaks. $T_{ij}^t$ and its estimate $ET_{ij}^t$ are almost identically distributed. The N2V measure of similarity, represented by the cosine distance in the vector space (Figure 1/c), is very close to being normally distributed in all cases. Finally, PPR, similar to the other TP measures, has a lognormal distribution with two peaks in the case of Astronomy & Astrophysics (Figure 1/d).

| Abbrev. | Definition |
|---|---|
| $T_{ij}^t$ | TP calculated using matrix operation. |
| $ET_{ij}^t$ | TP calculated using simulated random walks. |
| $ST_{ij}^t$ | Average shortest path TP. |
| SP | Shortest path distance. |
| N2V | Cosine similarity of nodes in embedding space generated with Node2vec. |
| PPR | Personalized page rank. |
| JAC | Jaccard similarity of backward and forward citations. |

***Table 2**. Tested metrics and their abbreviations*

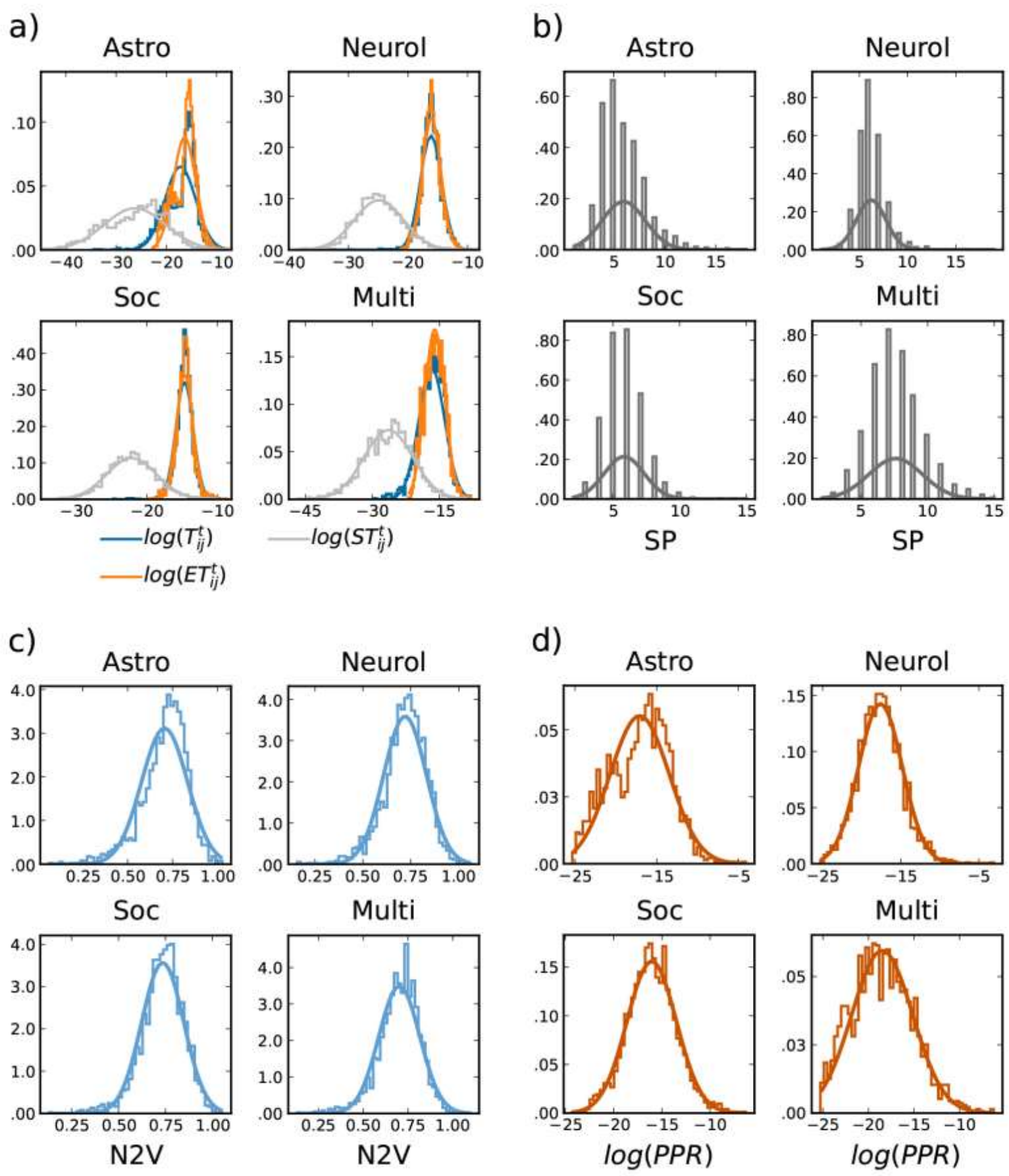

***Figure 1***. *Distribution of the measures with fitted normal curves.* ***(a)*** $T_{ij}^{t}$*,* $ET_{ij}^{t}$*, and* $ST_{ij}^{t}$ ***(b)*** *SP* ***(c)*** *N2V cosine distance* ***(d)*** *PPR.*

The direct implementation of the measurement concept is $T_{ij}^{t}$, and for this reason we examined how it relates to all the other similarity indices (Fig 2). Overall, its direct estimate $ET_{ij}^{t}$ has the strongest correlation with $T_{ij}^{t}$ (Fig 2/a). The next strongest association of $T_{ij}^{t}$ is with $ST_{ij}^{t}$ and PPR (Fig 2/a and Fig 2/d). The relationship of $T_{ij}^{t}$ and $ST_{ij}^{t}$ is not perfectly linear, but $ST_{ij}^{t}$ gives consistently good approximations for all ranges of $T_{ij}^{t}$ and in all fields. SP has a weaker correlation with $T_{ij}^{t}$ (Fig 2/b) compared to the previous indices. As we mentioned above, we expected that N2V will correlate less with $T_{ij}^{t}$ at weaker similarity ranges. We fitted piecewise linear regression models (OLS) to see if N2V and $T_{ij}^{t}$ has differing magnitudes of association at higher and lower TP values (Fig 2/c). The regression lines were fitted above and under the median $T_{ij}^{t}$. These slopes are considerably different; for high $T_{ij}^{t}$ values, the association is much stronger, and in most cases the association turns negative at weaker similarities. We produced similar results as Yun (2022) in relation to N2V and PPR. We explored the parameters of the embedding method to make sure that the non-linear and uneven association with $T_{ij}^{t}$ is not an idiosyncratic result. Our analysis can be found in Appendix B, and we can conclude this discrepancy is observable with all parameter configurations for N2V.

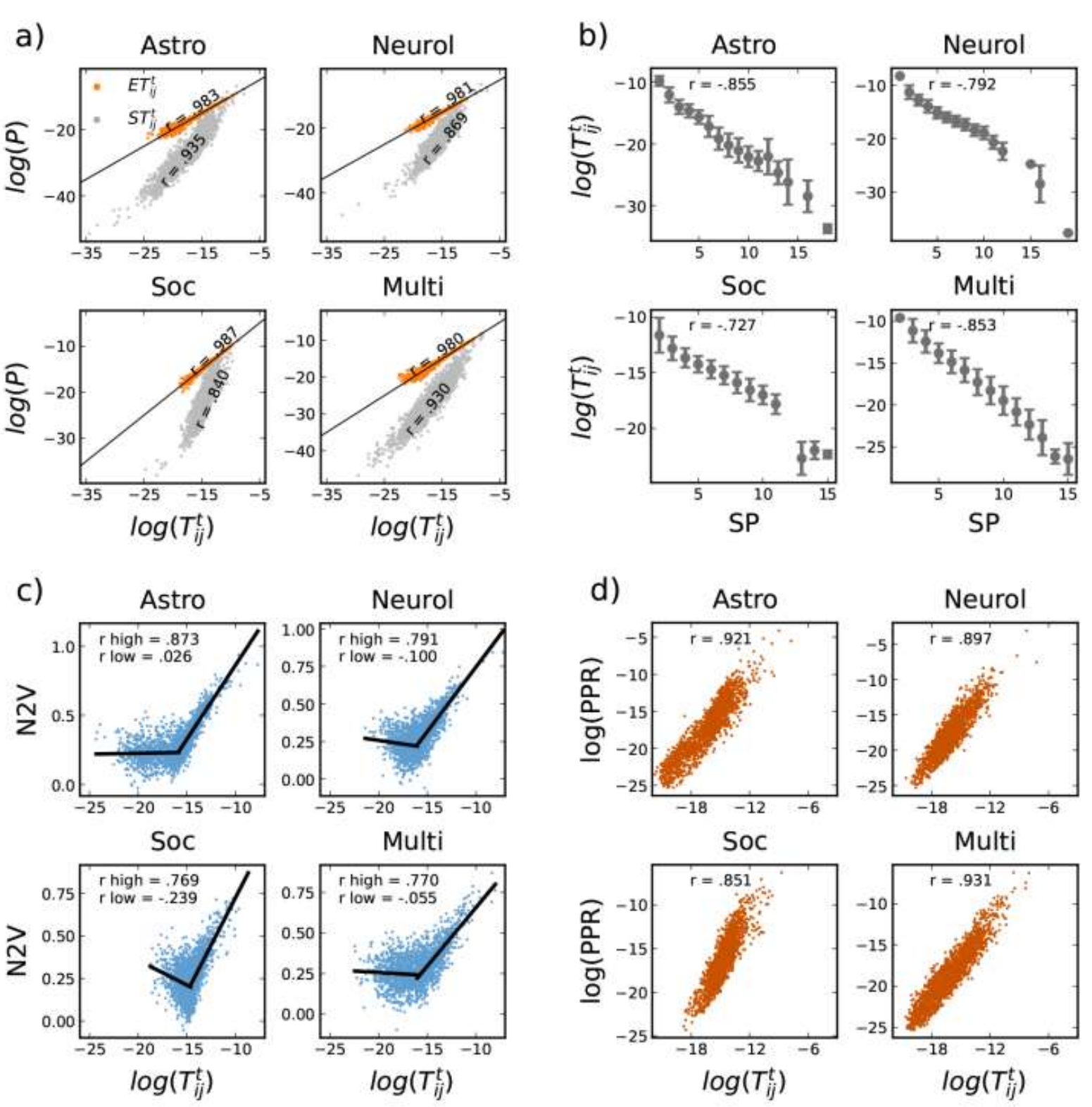

***Figure 2**. Bivariate distribution of $T_{ij}^{t}$ and the alternative measures. r stands for Pearson correlation. **(a)** $T_{ij}^{t}$, $ET_{ij}^{t}$, and $ST_{ij}^{t}$ **(b)** SP, means and standard deviations. **(c)** N2V cosine distance. The piecewise regression lines are fitted under and above the median of $T_{ij}^{t}$. **(d)** PPR.*

To compare the classification performance of the measures, we used the area under the curve score (AUC). Looking first at the co-authorship prediction (Figure 3), we can observe that, on average, the TP measures perform quite well, with AUC values around 0.85–0.9 in the best cases. However, in the case of clinical neurology, the average AUC drops to 0.74. Perhaps in this field, the clinical work setting mixes a diverse set of researchers, and the clinical setting may be a more important determinant of collaborative work than one's actual research specialty. However, this is merely speculation. Returning to the measurement evaluation, JAC is the worst performer, while SP is the second worst.

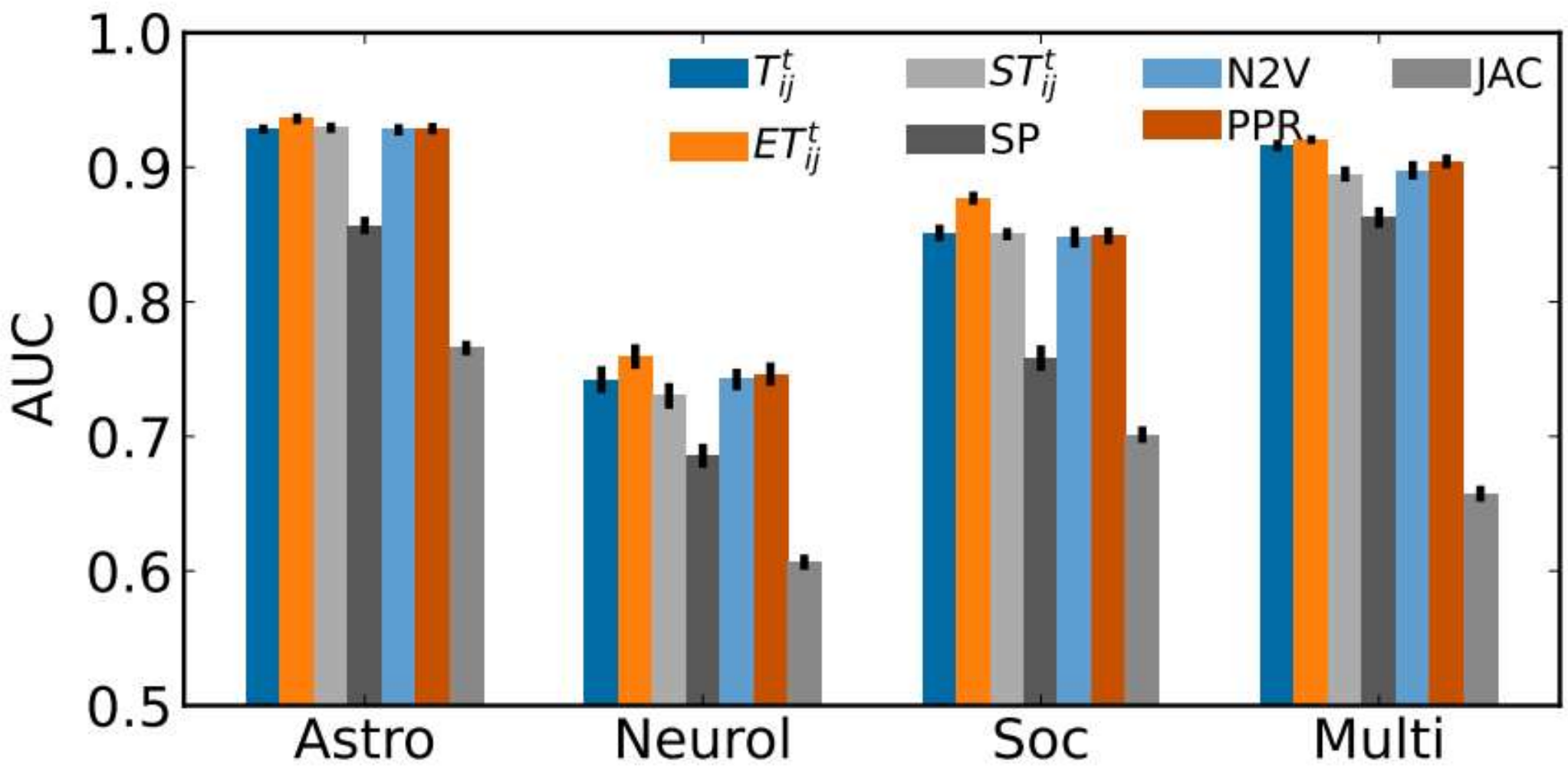

***Figure 3****. Co-author prediction AUC scores. Error bars are 95% confidence intervals form bootstrapping.*

Figure 4 presents the results for journal classification. The average AUC score drops from co-author prediction, and the performances spread is wider. The scores for the multidisciplinary dataset are at 0.5, suggesting that it is difficult to differentiate papers published in *Science* and *Nature*. The best performers are $T_{ij}^{t}$ and $ET_{ij}^{t}$ (0.73-0.79 AUC), followed by N2V (0.70-0.77 AUC). PPR significantly falls behind N2V in the case of neurology and sociology (0.66-0.70 AUC). SP and JAC are the worst performers, with JAC almost has no predictive capability, because at that similarity range reference and citation overlap is down to 1-2 percent.

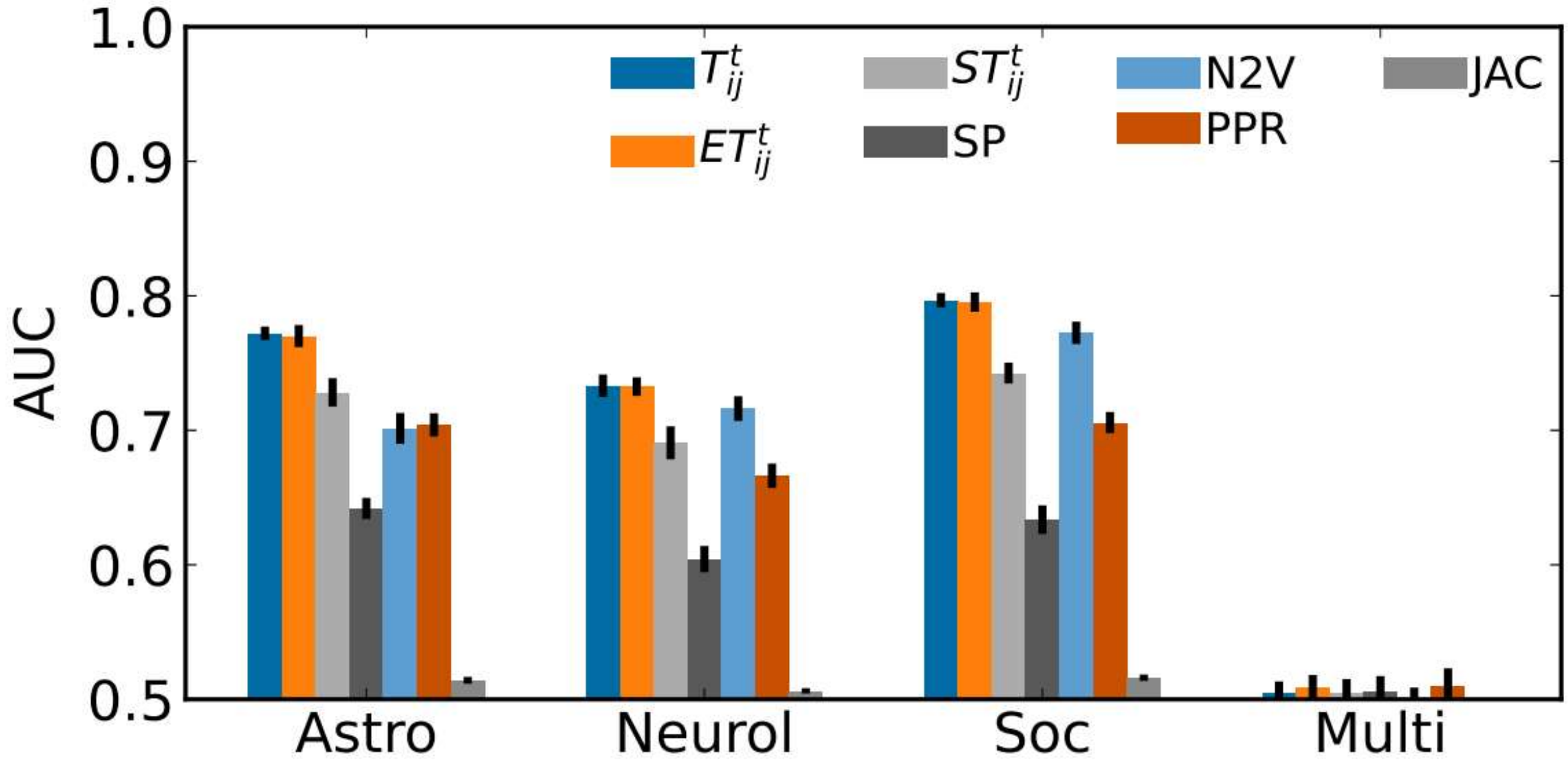

***Figure 4.*** *Journal classification AUC scores. Error bars are 95% confidence intervals form bootstrapping.*

The classification of disciplinary affiliations provides further insight into the utility of the studied measures (Figure 5/a). The main finding is that N2V significantly underperformed compared to all TP measures including PPR. While the best performers $T_{ij}^t$ and $ET_{ij}^t$ scored around 0.73, the AUC of N2V is 0.6. PPR and $ST_{ij}^t$ performed similarly to $T_{ij}^t$ and $ET_{ij}^t$ with AUC scores 0.71. SP came closest to the TP measures in this test, with a 0.67 AUC score. JAC is omitted from the plot, as almost all paper-pair values were zero.

We also present the results of an additional analysis of the macro disciplinary structure, specifically focusing on the performance of TP measures versus N2V. The purpose is to show further evidence that N2V underperforms at the macro level. We arrange the disciplines from life to physical sciences, starting with environmental sciences and ending with physics. Figures 5/b-c show similarity measures for each discipline in a matrix format for $T_{ij}^t$ and N2V. Based on the logic of the ordering, we expected that the similarities would be strongest around the diagonal of the resulting matrix, as any given discipline should be closest to its neighboring discipline along the continuum. This pattern is more noticeable in the case of $T_{ij}^t$, where the diagonal is more prominent. Agriculture, medicine, biology, and chemistry show an affinity with one another. In both figures, biology is relatively strongly related to all life sciences, though this is more evident in the case of $T_{ij}^{t}$.

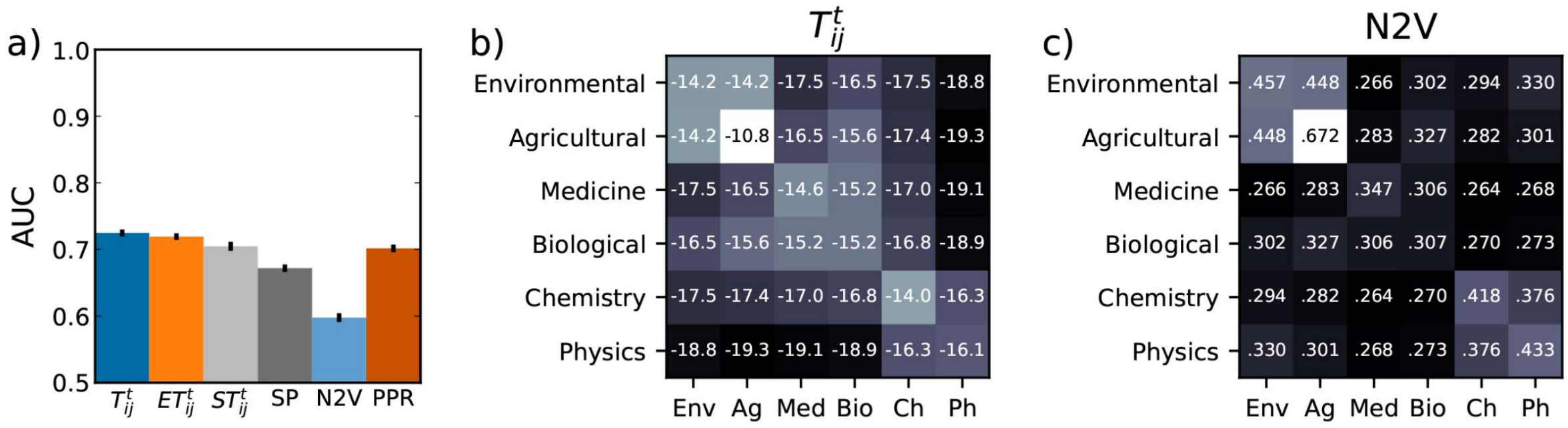

***Figure 5***. *Macro structure prediction and mapping.* ***(a)*** *Disciplinary co-classification AUC scores. Error bars are 95% confidence intervals form bootstrapping.* ***(b-c)*** *Discipline similarity matrices. The values in the cells are averages of the pertaining similarity measures. The shading is proportional to cell values.*

In section 3 we mentioned that the TP measures are related to the nodal degree. A random walk is more likely to pass through a high degree node. We evaluated this degree bias of the tested measures and showed the results on Figure 6. These results are based on the randomly sampled paper-pairs. Since PPR is asymmetric, we plotted the correlation coefficient for the nodes in the pair separately for each measure. N2V has a negligible correlation with degree. This desirable property of N2V is well known. JAC also has a small degree bias according to our test. The spearman correlation of $T_{ij}^{t}$ and $ET_{ij}^{t}$ with degree is moderate (0.05-0.19). $ST_{ij}^{t}$ has a slightly higher correlation ranging between 0.09 to 0.21. The degree bias of PPR has the same range 0.09 to 0.21 on the source node side of the dyad. However, the target node's degree has a strong correlation with PPR. This is not a surprise given that PageRank was introduced originally as a webpage popularity ranking algorithm. Finally, shortest paths are also strongly dependent on the degree.

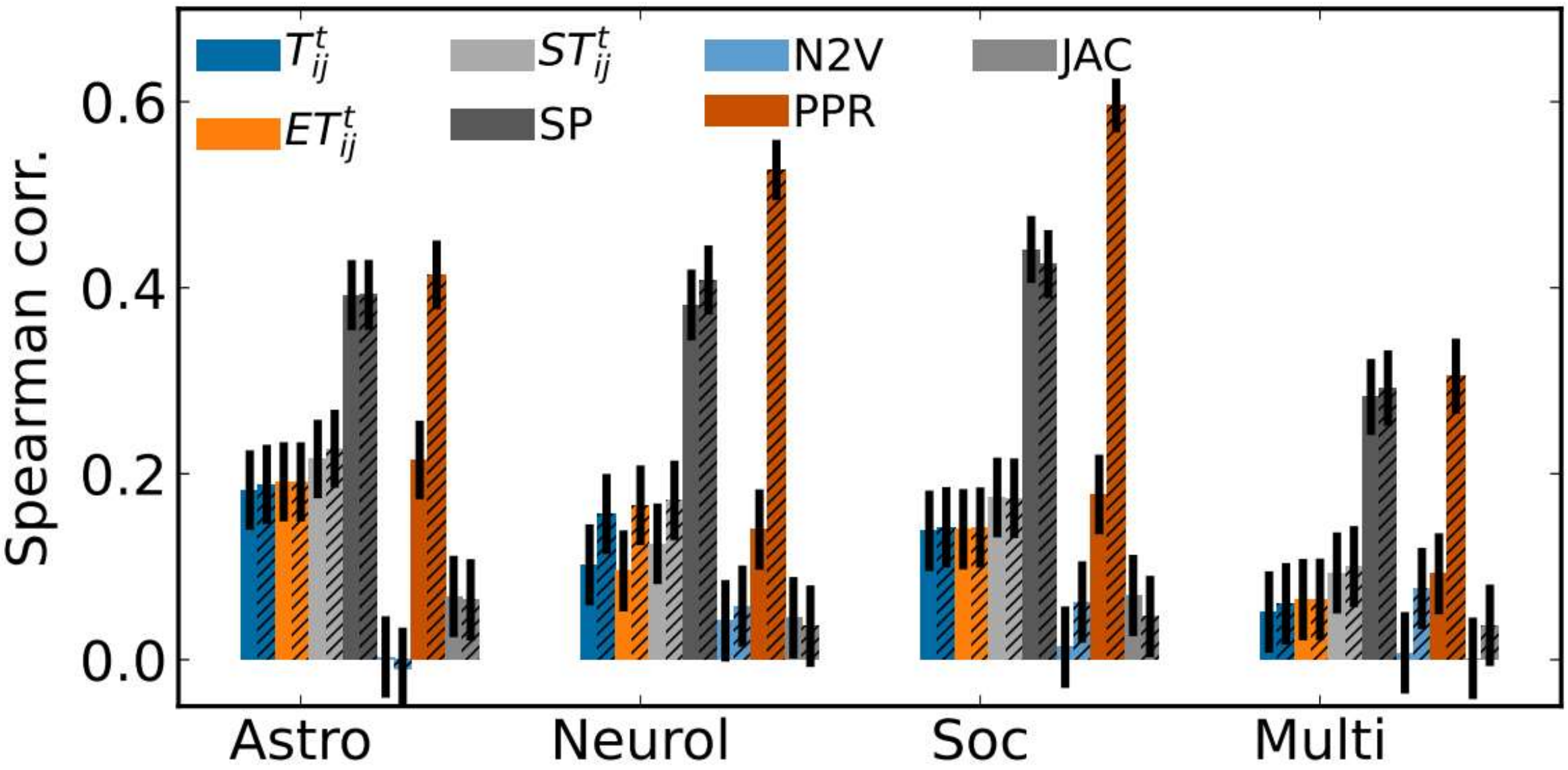

***Figure 6.*** *Degree bias of the seven similarity measures. Error bars are 95% confidence intervals of the Spearman correlation coefficient. The plain bars represent the source nodes, and the patterned bars indicate the target node's correlation coefficient.*

The evaluation of the runtimes is shown in Figure 7. These tests were conducted on random networks generated using the stochastic block model (Karrer & Newman, 2011). Such networks replicate the underlying modular structure commonly observed in real-world networks, where nodes are grouped into communities (such as disciplines), with a higher probability of connections within the same group and a lower probability between different groups. This approach allows us to evaluate the performance of our methods under controlled conditions, simulating community structures found in scholarly networks. Our results indicate that the speed for $ET^t_{ij}$ scales best with network size, with $T^t_{ij}$ feasible only for small networks in practice. Overall PPR calculated with the Power Iteration with Forward Push method appears to be the slowest. However, the computational speed somewhat stabilizes for larger networks. The employed parameter setting contributes to the method's slowness. This parameter setting (Appendix A) is necessary to achieve a sufficient level of precision for the validation tests. These tests were demanding for PPR because the probability that the walker restarting had to be set very low, so that the walkers could reach beyond the immediate neighborhoods of the focal nodes.

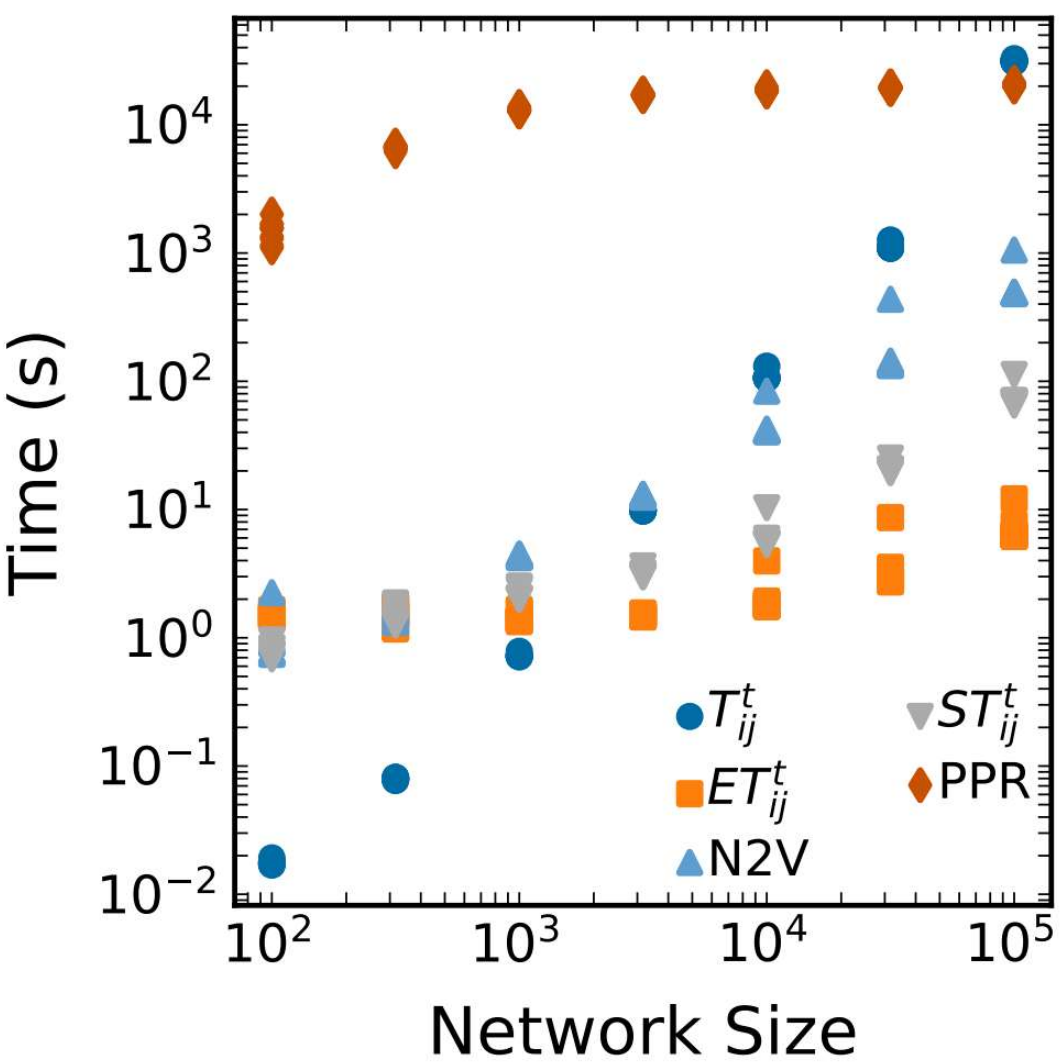

***Figure** 7. Runtime benchmarks on random networks. 10 random networks for each network size and method.*

## 8. Conclusions

We presented an interpretable, continuous, and symmetrical family of measures for paper and author similarity. The definition of these measures has explicit conceptual foundations, which leads straightforwardly to their operational definition based on random walk TPs. We found that our main metric $T_{ij}^t$ and its direct estimate $ET_{ij}^t$ are the most consistent measures of similarity across different scales of analysis (co-authorships, journals, and disciplines), outperforming our other proposed solutions as well as rival similarity measures. Additionally, $ET_{ij}^t$ was found to be reasonably lightweight to implement. The closest alternative to the concept behind $T_{ij}^t$ and $ET_{ij}^t$ is PPR, which prioritize local similarity precision over long distance similarity exploration. This explains why it performed slightly less effectively despite considerable computational investment.

The major limitation of our approach stems from the fact that publication data does not represent equally well the research interest of scientists across different fields of inquiry. Moreover, our metrics exhibit some level of degree bias: high degree nodes (papers) appear to be more similar to other nodes in the citation network. This is an undesirable quality for a similarity measure. However, we want to point out that by moving higher degree papers closer, $T_{ij}^t$ remains consistent with its conceptual definition: high degree papers are simply easy to reach, while someone browsing the literature. In other words, highly cited papers are more visible and more likely to bridge authors and fields. Finally, we want to avoid the impression that we aim to measure the actual interests of researchers. Rather, our terminology expresses the interests of a generic information searching agent without a concern about the specific content of the browsed literature. The logic of this approach is to map the enabling and constraining structure of information encoded in the citation network to trace similarity. However, as we expressed in the introduction, the measurement approach can be developed further to include nuances that represent more closely some specific search profiles.

Choosing the appropriate walk length for $T_{ij}^{t}$ and $ET_{ij}^{t}$ is an important step. If the network is very sparse and/or highly clustered, zero similarity values could be an issue. In this case one should increase the walk length. Given the small world property of most networks, the shortest path length between two nodes should be relatively short. A relevant property of TP – related to the previous discussion on degree correlation – is that as the walk length $\tau$ increases, the TP defined as $(P^{\tau})_{ij}$ converges to the degree of $j$. However, $\tau$ must be very large to induce this effect in large networks. Overall, we did not find reason to go beyond walk lengths over approximately 20, as this would undesirably slow down computation.

We have two observations about relying on paper-level information to represent author level-behavior. Both observations highlight a general issue stemming from the loose relationship between authorship status and author contributions. These issues become more pronounced as the prevalence of multiple authors increases. First, in our experience, only considering papers where the focal author is in the first or last authorship position serves as a better predictor of future collaborators. We used this approach across all disciplines, but it may be beneficial to calibrate this step for certain disciplinary publishing cultures. The main disadvantage of this choice is information loss, as it becomes harder to assess an author's interests if they have not published enough papers as a first or last author. The second observation is that, in some fields, publications represent researchers' true interests less accurately. As we already mentioned, the usefulness of paper-level data to represent author interests is inherently limited. This issue was particularly evident in clinical neurology, where we speculated that the work setting of clinical research might explain the discrepancy. It is possible that substantive research interests and clinical practice do not always align when researchers collaborate. Perhaps data collection and specialized expertise diverge in clinical research, although both contribution types qualify for authorship.

**Acknowledgements**: The authors thank the support from the Air Force Office of Scientific Research under Award #FA9550-19-1-0391. This work used JetStream2 at Indiana University through allocation CIS230183 from the Advanced Cyberinfrastructure Coordination Ecosystem: Services & Support (ACCESS) program, which is supported by National Science Foundation grants #2138259, #2138286, #2138307, #2137603, and #2138296.

## Appendices

***Appendix A:*** *Parametrization of the tested measures*

The walk length $t$ is set to 20 for all analysis, and the number of walkers to test TP is 1 million from each direction ($i-j$ and $j-i$). The teleportation probability for PPR estimation (that is the probability that the walker restarts) is .4, and the tolerance value for the forward push algorithm is 1e-11. The window length for N2V is 10, the space has 64 dimensions, and the walk length is 80. These parameters have been chosen to provide feasible and comparable accuracy for each method given the data sizes. By taking random samples of dyads with a size of 2000 paper-pairs in each discipline we found that all paper-pairs are reachable through at least one path within 20 steps. The proportion of zero $ET_{ij}^{t}$ and PPR estimates range between 14 to 0.4 percent. With these parameter settings the discrepancy of missing estimates between $ET_{ij}^{t}$ and PPR is negligible: it is around 0.1-0.7 percent across the different datasets. We experimented with several parameter configurations for N2V and chose one with a goof fit to $T_{ij}^{t}$ (see the Appendix for details).

***Appendix B:*** *N2V parameter exploration*

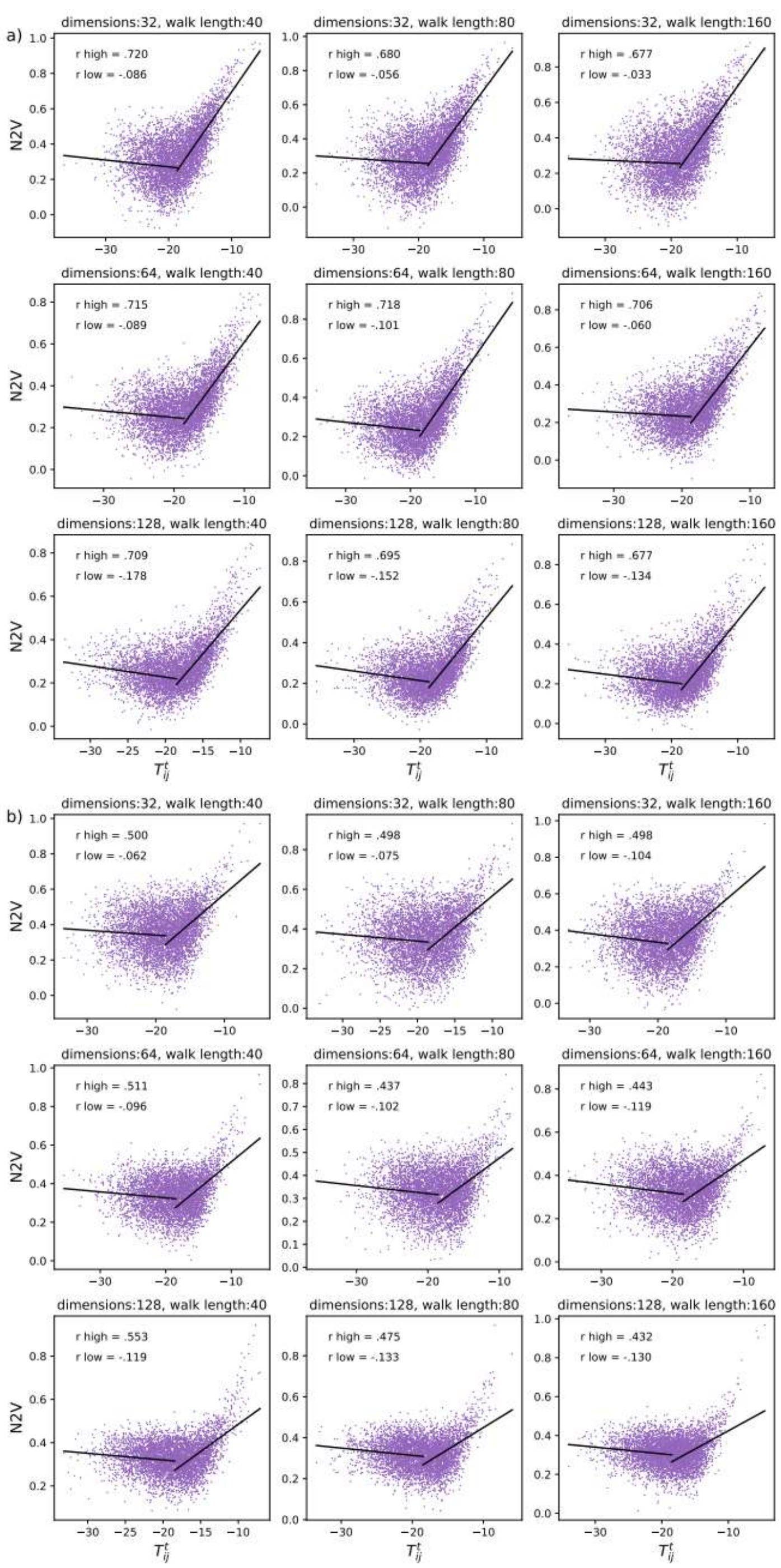

***Appendix Figure 1.*** *Testing parameters of N2V procedure, performed on the multidisciplinary dataset. The three manipulated parameters are window (or context) length (2, 10), dimensions (32, 64, 128), and walk length (40, 80, 120). TP stands for* $T_{ij}^t$. ***(a)*** *Window = 10.* ***(b)*** *Window = 2.*

***Appendix C:*** *The TP package*

We developed a Python package to calculate the similarity metrics used in this study, including $T_{ij}^{t}$, $ET_{ij}^{t}$, N2V, SP, and $ST_{ij}^{t}$. The package is available via PyPI, and the source code can be accessed on GitHub at https://github.com/filipinascimento/tpsimilarity. It features a random walk sampler implemented in C, enabling fast $ET_{ij}^{t}$ calculations even for large networks.